\pdfoutput=1
\documentclass[conference]{IEEEtran}
\ifCLASSINFOpdf
  \usepackage[pdftex]{graphicx}
\else
  \usepackage[dvips]{graphicx}
\fi
\graphicspath{{./figures/}}
\usepackage{amsmath}
 \usepackage{gensymb}

\usepackage[utf8x]{inputenc}

\ifCLASSOPTIONcompsoc
  \usepackage[caption=false,font=normalsize,labelfont=sf,textfont=sf]{subfig}
\else
  \usepackage[caption=false,font=footnotesize]{subfig}
\fi
\begin{document}
%
\title{Foveated Video Streaming for Cloud Gaming}
%

\author{\IEEEauthorblockN{Gazi Illahi, Matti Siekkinen}
\IEEEauthorblockA{Dept. of Computer Science, School of Science\\
Aalto University, Finland\\
Email: \{gazi.illahi,matti.siekkinen\}@aalto.fi}
\and
\IEEEauthorblockN{Enrico Masala}
\IEEEauthorblockA{Control and Computer Eng. Dept.\\
Politecnico di Torino, Italy\\
Email: enrico.masala@polito.it}}


%


\maketitle
\begin{abstract}
Good user experience with interactive cloud-based multimedia applications, such as cloud gaming and cloud-based VR, requires low end-to-end latency and large amounts of downstream network bandwidth at the same time. In this paper, we present a foveated video streaming system for cloud gaming. The system adapts video stream quality by adjusting the encoding parameters on the fly to match the player's gaze position. We conduct measurements with a prototype that we developed for a cloud gaming system in conjunction with eye tracker hardware. Evaluation results suggest that such foveated streaming can reduce bandwidth requirements by even more than 50\% depending on parametrization of the foveated video coding and that it is feasible from the latency perspective.
\end{abstract}


%
\IEEEpeerreviewmaketitle

\section{Introduction}
\label{sec:intro}

There is an increasing class of interactive applications that require high-end graphics rendering. In some cases, remote rendering of the graphics is convenient or even necessary due to limited local computing resources. A good example is cloud gaming in which the game logic is also executed remotely and a thin client software simply forwards control commands, such as key press events, to the remote server, and decodes a video stream encoded and transmitted by the server~\cite{shea2013cloud}. Other emerging examples are mobile Virtual Reality~\cite{boos16flashback} and real-time 360\degree video~\cite{silva16siggraph}.

It is clear that both low end-to-end latency from motion or key press to photon and a high quality video stream are imperative for good user experience with these applications. For this reason, they impose challenging bandwidth and latency requirements on the network infrastructure and the problem gets aggravated when they are used with wireless mobile network connectivity. In this paper, we address the bandwidth challenge by investigating the use of so called \textit{foveated video streaming}. Fovea is the part of a human eye responsible for the sharp central vision and by foveated streaming we mean that the quality or resolution of the video is adjusted based on user's gaze tracking information so that the region around the center of the gaze position is encoded with high quality or resolution and the peripheral regions with low quality or resolution. As a result, significant bandwidth savings can be achieved without noticeable degradation in quality of experience.

Foveated image and video coding has been studied for a relatively long time. However, it has only recently drawn renewed interest because of affordable gaze tracking solutions, new applications scenarios, and technological evolution, especially in the networking domain, enabling low enough latency. Lately several papers have been published on foveated streaming but they have mainly focused on pre-encoded content and using video tiling. Our objective in this paper is to explore the feasibility of a real-time foveated streaming solution for cloud gaming. The scheme we use is based on dynamic adjustment of the quantization parameters of the video encoder and it requires minimal amount of modifications to existing cloud gaming software. 

We have developed a prototype by integrating together an off-the-shelf external eye tracking device with a state of the art cloud gaming software. Our evaluation results suggest that foveated streaming can potentially reduce bandwidth consumption dramatically. The exact effect depends on how the scheme is parametrized but does not seem to depend much on the game type, even though we discover clear differences in gaze movement patterns between the different games we studied. Based on our own subjective experience as well as some back of the envelope calculations of latency combined with gaze data analysis from gaming sessions with different games, we are rather optimistic that latency achievable with current technology is short enough for good user experience.

\section{Background}
\label{sec:bg}

\subsection{Cloud Gaming}
\label{sec:cloud_gaming}

\begin{figure}[ht]
 \begin{center}
   \includegraphics[width=0.7\linewidth]{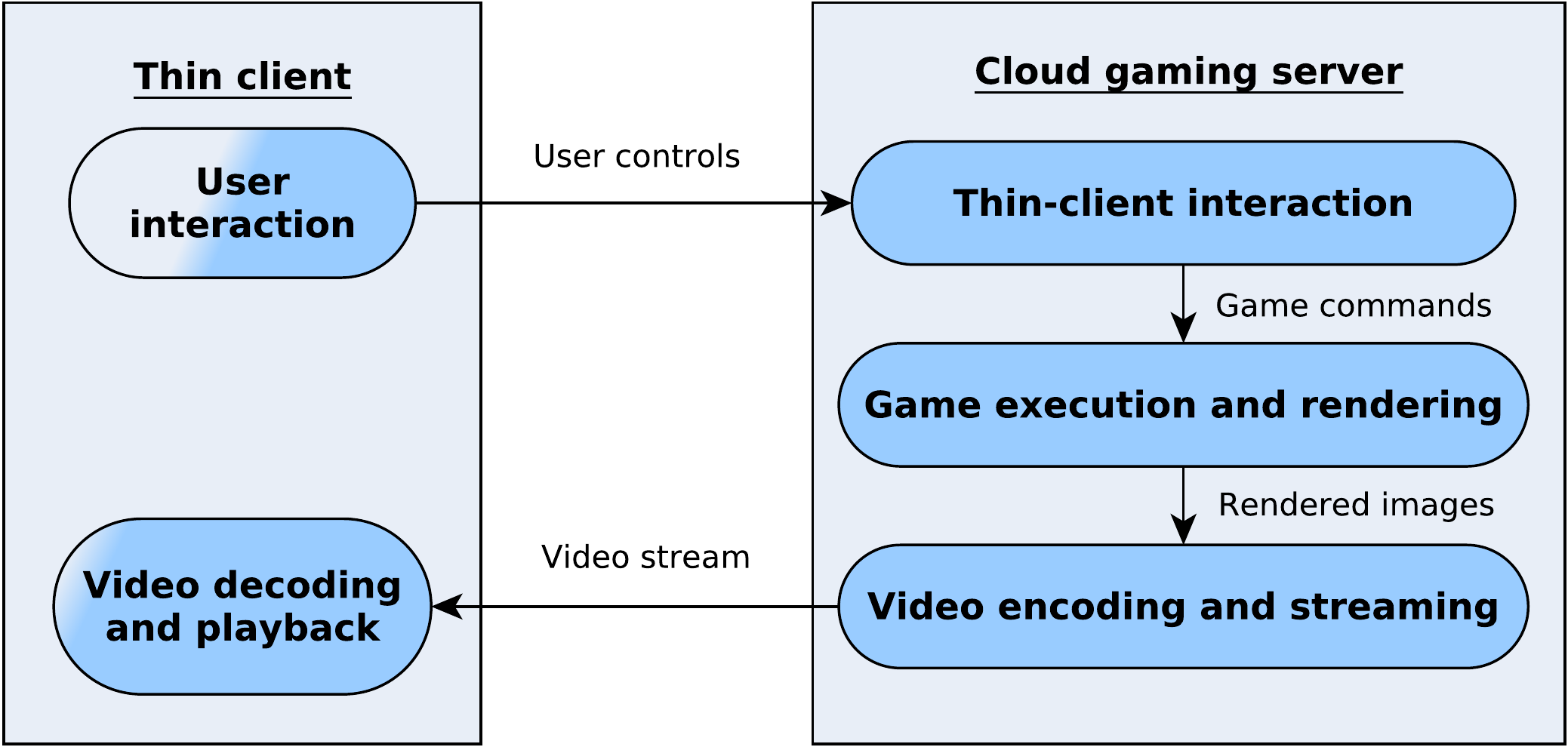}
   \caption{Cloud gaming architecture~\cite{kamarainen17mmsys}.}
   \label{fig:cloud_gaming}
 \end{center}
\end{figure}

Cloud gaming system consists of a thin client and a remote server as illustrated by Figure~\ref{fig:cloud_gaming}~\cite{kamarainen17mmsys,shea2013cloud}. The server is typically a virtual machine residing in a cloud. The client software forwards all control events from the user to the server which runs the game as usual and renders all the graphics. The cloud gaming software at the server side replays the user controls locally and produces a video stream of the graphics either by compressing video captured from the screen or by obtaining the compressed video frames directly from the graphics card. The client software decodes the video stream and displays it on screen. This approach does not require any modifications to regular PC games to be run as cloud games and allows graphically demanding games that require heavy GPU processing to be played with low end client devices, even mobile devices.

The whole process should be as invisible to the user as possible, which means that the end-to-end latency, i.e. delay from the moment user issues control input to the moment when the user perceives the resulting action on display, should be preferably shorter than 100 ms~\cite{kamarainen17mmsys} and the available amount of downstream bandwidth sufficiently high so that high quality video stream can be delivered.

\subsection{Foveated Video Coding}
\label{sec:foveated_video_coding}

A good real-time streaming system requires an encoder that can provide the best quality while fulfilling some constraints, typically dictated by the available bandwidth. The usual way that video streaming services cope with bandwidth constraints today is to adapt the overall target video bitrate according to measured or estimated amount of available bandwidth. Foveated video coding can be viewed as a complementary solution that reduces the overall bandwidth requirements by allocating more bits to the region of the image currently being actively observed by the user.
Therefore, such a region can have better quality with respect to other regions that are currently less important for the observer.

The key parameter allowing to change the tradeoff between compression and quality in video coding is the quantization parameter (QP). Such a value determines
how well details are reconstructed after the quantization process. The lower the value, the better the quality and vice versa.
The QP value can be adjusted for each single macroblock in which the image is subdivided by the video encoder.
Any video encoder provides its own set of strategies to control such a value so that the desired rate quality tradeoff can be achieved.
Strategies may vary from extremely simple ones, such as constant QP, to rate-distortion optimized ones, where different values of the QP and other coding parameters are
explored during the encoding process so that the best quality is achieved for the given rate quality tradeoff.

In this work we employ the AVC standard~\cite{h264:standard} for video encoding by using the {\em x264} software implementation. Such an encoder allows to use different strategies to control
the rate quality tradeoff. We focus on the so-called constant rate factor (CRF), which is a variant of the constant QP encoding, in which each
frame is compressed by a different amount so to achieve a certain level of perceived quality. The different sensitivity of the human eye to details for
still and moving objects is exploited by the CRF scheme to save bits, i.e., drop more details, depending on the amount of motion in the frame.
This is currently regarded as the best encoding mode for the {\em x264} software when only one pass encoding is possible, as for the case of encoding for streaming with
very low latency.

In order to achieve foveated video coding, we used the input from the gaze tracking system to adjust the QP value computed by the CRF algorithm
so that quality is reduced to the areas outside of user's current gaze position by increasing the QP for the remaining part of the frame.
In practice, an offset has been added to the QP value computed by the CRF algorithm depending on the position of the macroblock with respect to the gaze location. We describe the prototype implementation in more detail in the next section.

\section{Cloud Gaming with Foveated Streaming}
\label{sec:system}
 
\subsection{Cloud Gaming Software}

We use GamingAnywhere~\cite{Huang2013ga} as our cloud gaming platform. GamingAnywhere is an open-source portable cloud gaming software with extensible modular architecture, providing cloud gaming client and server software for various operating systems. The cloud gaming prototype we have built comprises a modified version of the server installed on a Linux machine and a client installed on a Windows machine connected to each other on a GbE local network. The GamingAnwhere server is capable of capturing game video, streaming it to a client, and relaying user control input to the game as received from the client. The GamingAnywhere client captures user input and sends it to the server and receives, decodes, and displays the video stream sent back by the server.
 
In our prototype, we modify the server to receive gaze location data from the client and use it for foveated encoding of the video rendered by the game application. The server is configured to use the {\em x264} encoder with a preset CRF and adaptive quantization enabled to allow usage of quantization offsets on a per macroblock basis. A Tobii 4C~\cite{Tobii} gaze tracker is installed on the client machine and the client is configured to transmit gaze location data of the user to the server as soon as the gaze tracker provides it.
 
\subsection{Gaze Tracker}
\label{sec:gaze_tracker}
The Tobii 4C eye tracker is a consumer grade eye tracking device directed towards gaming and interactive applications. It has an on board eye tracking ASIC and is capable of providing individual eye, gaze, gaze fixation and head location data. We install the Eye Tracker on the client machine and configure it to output lightly filtered gaze data. The light filtering smoothes out sudden eye movements, such as saccadic movements that are very fast eye movements jumping around the target and happen during certain phases of fixation, and noise based on current and past data points as well as velocity of eye movements ~\cite{TobbiDevGuide}. The client transmits the gaze position data to the server as soon as it is available from the gaze tracker. The eye tracker is also capable of providing fixation information but we decided to use the periodically sampled gaze data because we do not know how Tobii computes fixations. 

\subsection{Encoder-Gaze Tracker Interfacing}
\label{sec:encoder_tracker_interface}
\begin{figure}[t]
\centering
\subfloat[Visual acuity of human eye]{\label{fig:visual}\includegraphics[width=0.5\linewidth]{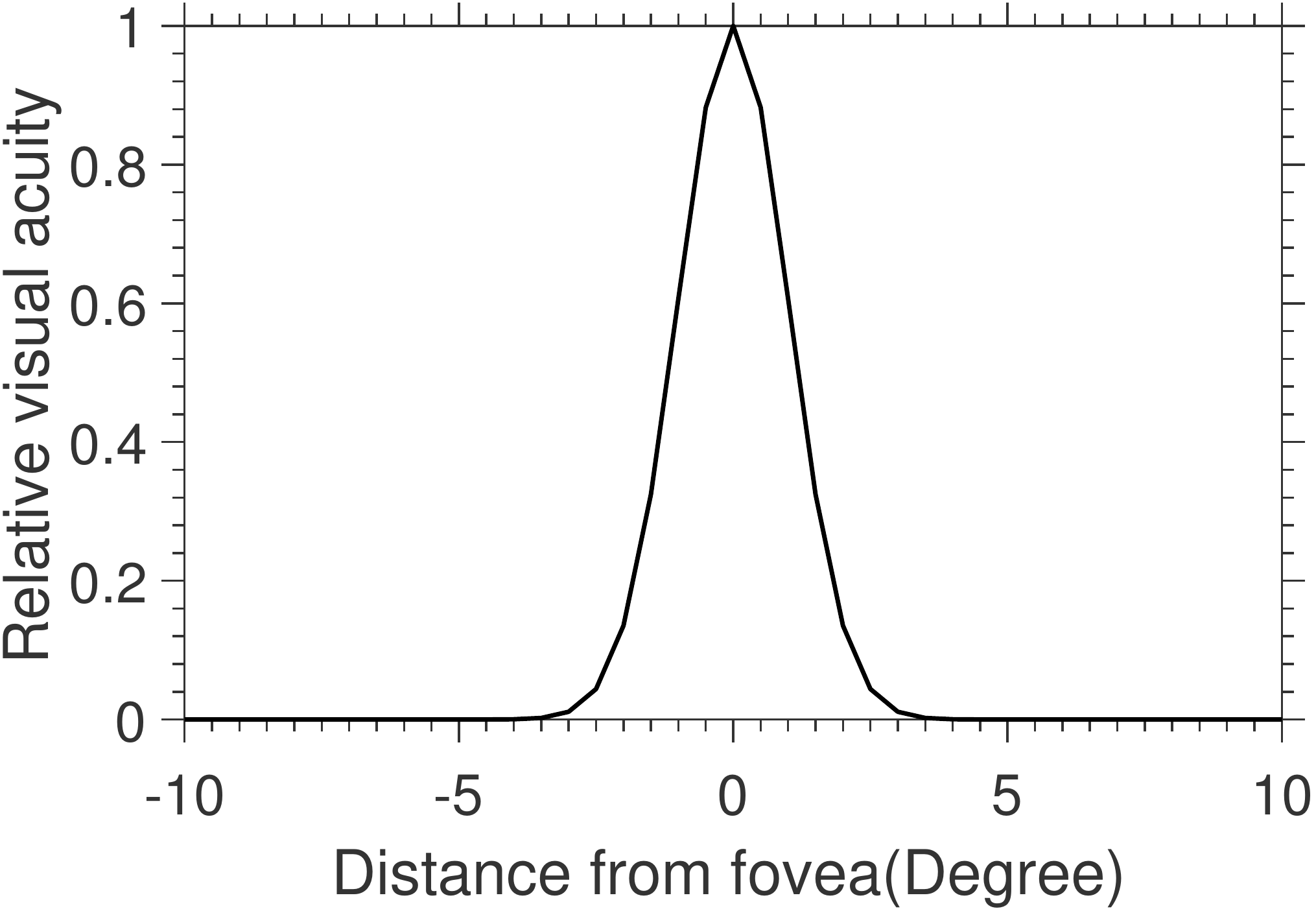}}
\subfloat[$QO$ at different values of $W$]{\label{fig:foveation}\includegraphics[width=0.5\linewidth]{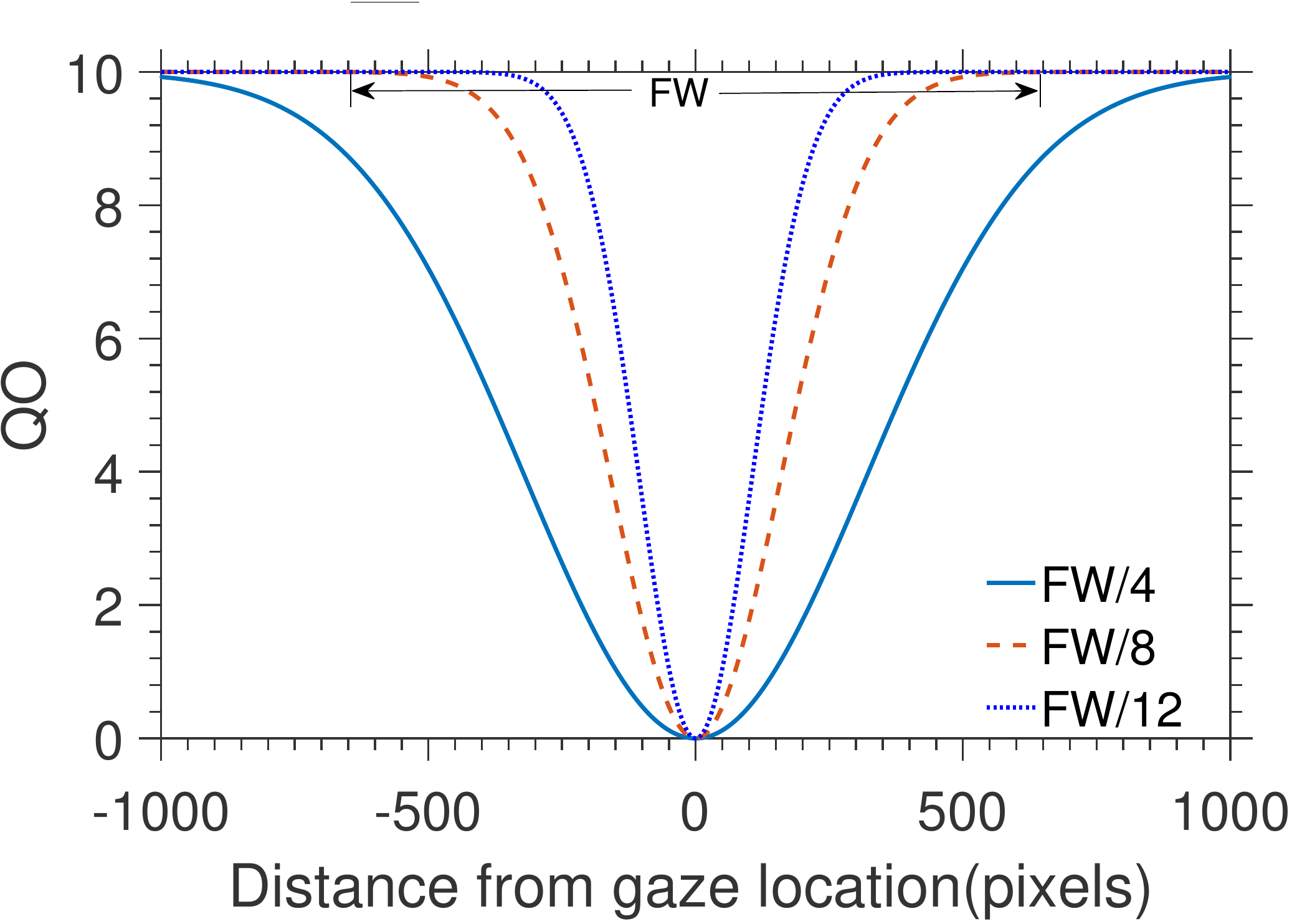}}
\caption{Foveation and $QO$ calculation. $FW$ is the width of the output frame in pixels}
\vspace{-5mm}
\end{figure}
 The density of cone cells of the human eye is high at the fovea and drops off rapidly farther from it~\cite{wandell}. Due to this, the relative visual acuity of the human eye also drops dramatically with the angular distance from the fovea. The fovea is believed to occupy only  2\textsuperscript{o} of the human visual field~\cite{ryoo16mmsys}, as illustrated by Figure \ref{fig:visual}. The server in our prototype is configured to encode each frame with a quality that reflects the relative visual acuity of the eye. 
 
In the server, the module feeding video rendered by the game logic to the encoder is configured to receive a gaze location data stream from the client. The received gaze location data is used to define the quantization parameters of each macroblock of the game video frame currently being encoded. The macroblocks being looked at by the player on the client side, as indicated by the gaze location, are encoded with the smallest quantization parameter values, while the macroblocks away from the gaze location are encoded with higher quantization parameter values. To this end, we use a QP offset array to define values to be added to QP values decided by the encoder's own algorithms for each macroblock. 

In a given video frame, the offset, denoted by $QO(i,j)$, for each macroblock indexed by $i,j$ is computed using a two dimensional Gaussian function according to \eqref{eq:offsets}.
\begin{equation}
QO(i,j) = QO_{max}\left(1-\exp(\frac{(i-x)^2+(j-y)^2}{2(W)^2})\right)
\label{eq:offsets}
\end{equation}
In \eqref{eq:offsets}, $QO_{max}$ is the maximum offset, $i$ and $j$ are indices of the matrix of macroblocks comprising the frame, $x$ and $y$ are the indices of the macroblock where the center of the gaze is, and $W$ controls the width of the foveal region. We define foveal region as the region on the screen of the client machine where the game video quality should be high. The parameters $QO_{max}$ and $W$ are tunable and allow us to investigate and evaluate the setup. The effect of $W$ on (normalized) offsets  $QO$ applied to macroblocks of a frame with respect to distance from gaze location is illustrated in Figure \ref{fig:foveation}. The size of the foveal region, that is the region which the human eyes perceives with highest visual acuity, depends on the viewing distance. The area becomes larger with increase in viewing distance as the area covered by the angle occupied by fovea in the human visual field increases with distance from the fovea. It should be noted that $W$ controls, rather than is equal to, the width of the foveal region, since we calculate $QO(i,j)$ based on a Gaussian curve with smoothly increasing values between 0 and $QO_{max}$ rather than as a step function with  demarcated region of $QO$=0 and $QO=QO_{max}$. We believe this follows the visual acuity of human eye more naturally and also ameliorates, to some extent, any inaccuracy in gaze tracking.

\section{Evaluation}
\label{sec:eval}

\subsection{Measurement Setup}

To investigate the effect of foveated video encoding, we setup our cloud gaming system prototype and performed measurements while playing four different kinds of games: AssaultCube is a First Person Shooter (FPS) game, Trine 2 is a side-scrolling adventure game, Little Racers STREET (we abbreviate it as Little Racers) is a car racing game from bird's-eye perspective, and Formula Fusion is a futuristic anti-gravity racing game with a viewpoint from behind the vehicle. The client and server were deployed within the same GbE network having negligible latency in order to avoid any network related bias to the results.

For traffic measurements, we captured  all traffic flowing between the cloud gaming server and client using \texttt{tcpdump} ~\cite{tcpdump}. Throughput per second was extracted from the captured data using Wireshark~\cite{wireshark}. A set of measurements for a game consisted of same gameplay over the same period of time, effort being made to replicate the player actions in each measurement as much as possible. The x264 encoder is set with the following parameters based, in part, on the recommendations in ~\cite{Huang2013ga}:\\
\texttt{--profile main --preset ultrafast --tune zerolatency
--crf 28 --ref 1 --me\_method dia --me\_range 16 --keyint 48
--intra-refresh   --threads 4}

\subsection{Bandwidth Savings}

We first study the potential of foveated streaming to reduce the bandwidth requirements in cloud gaming. To this end, we performed a series of measurements in which we either vary the maximum offset $QO_{max}$ or the $W$ parameter that controls the foveal region while keeping the other parameter constant. The values of $W$ are defined relative to the frame width of output video ($FW$) in pixels. Defining $W$ relative to $FW$ is a simple solution of providing screen size agnostic visual quality. Furthermore, this solution scales well with the viewing distance as typically the viewing distance increases with size of the display screen, thereby increasing the width of the area of high visual acuity.

\begin{figure}[t]
\begin{center}
\subfloat[AssaultCube]{\includegraphics[width=0.9\linewidth]{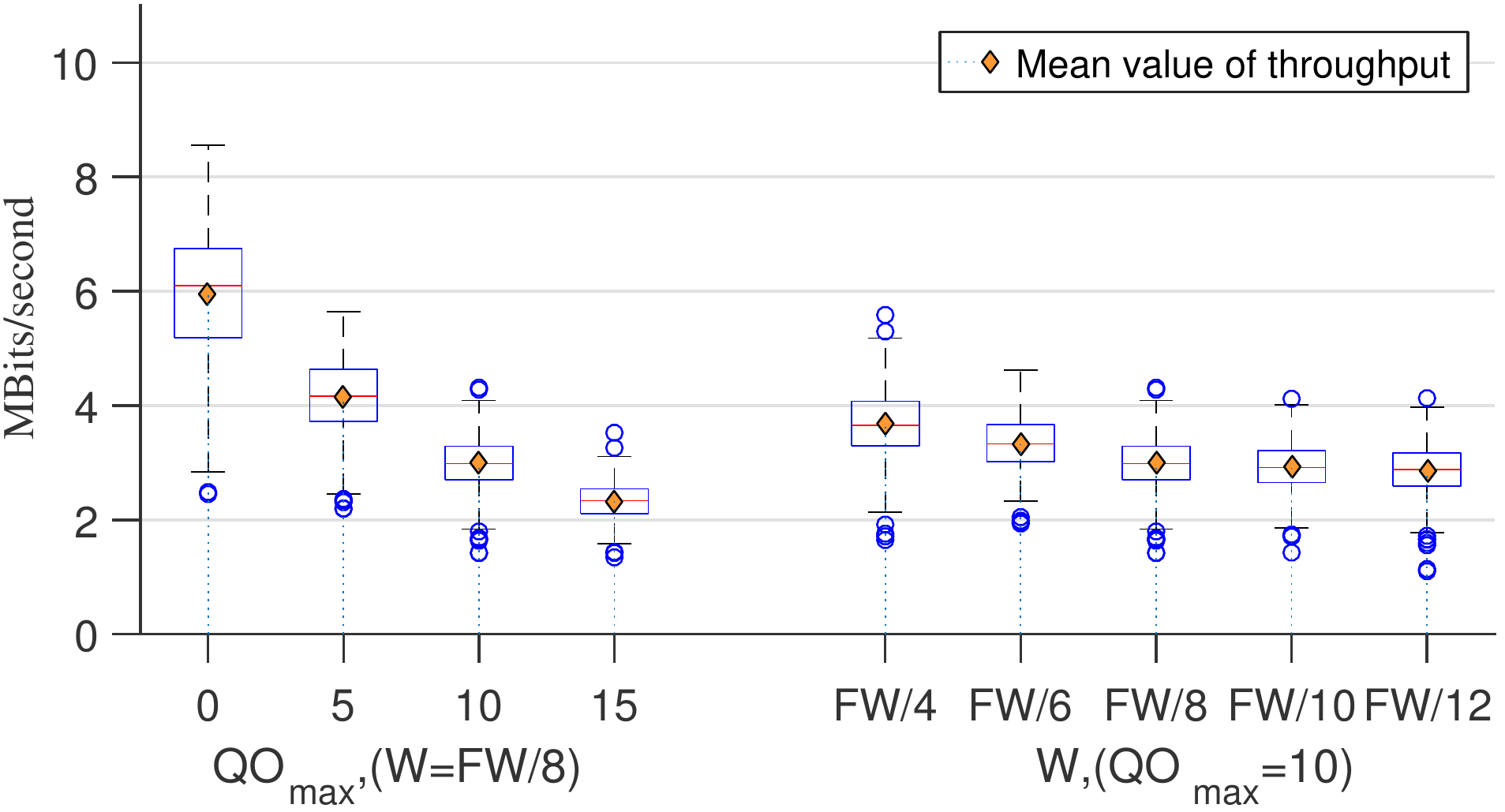}}\\
\subfloat[Trine 2]{\includegraphics[width=0.9\linewidth]{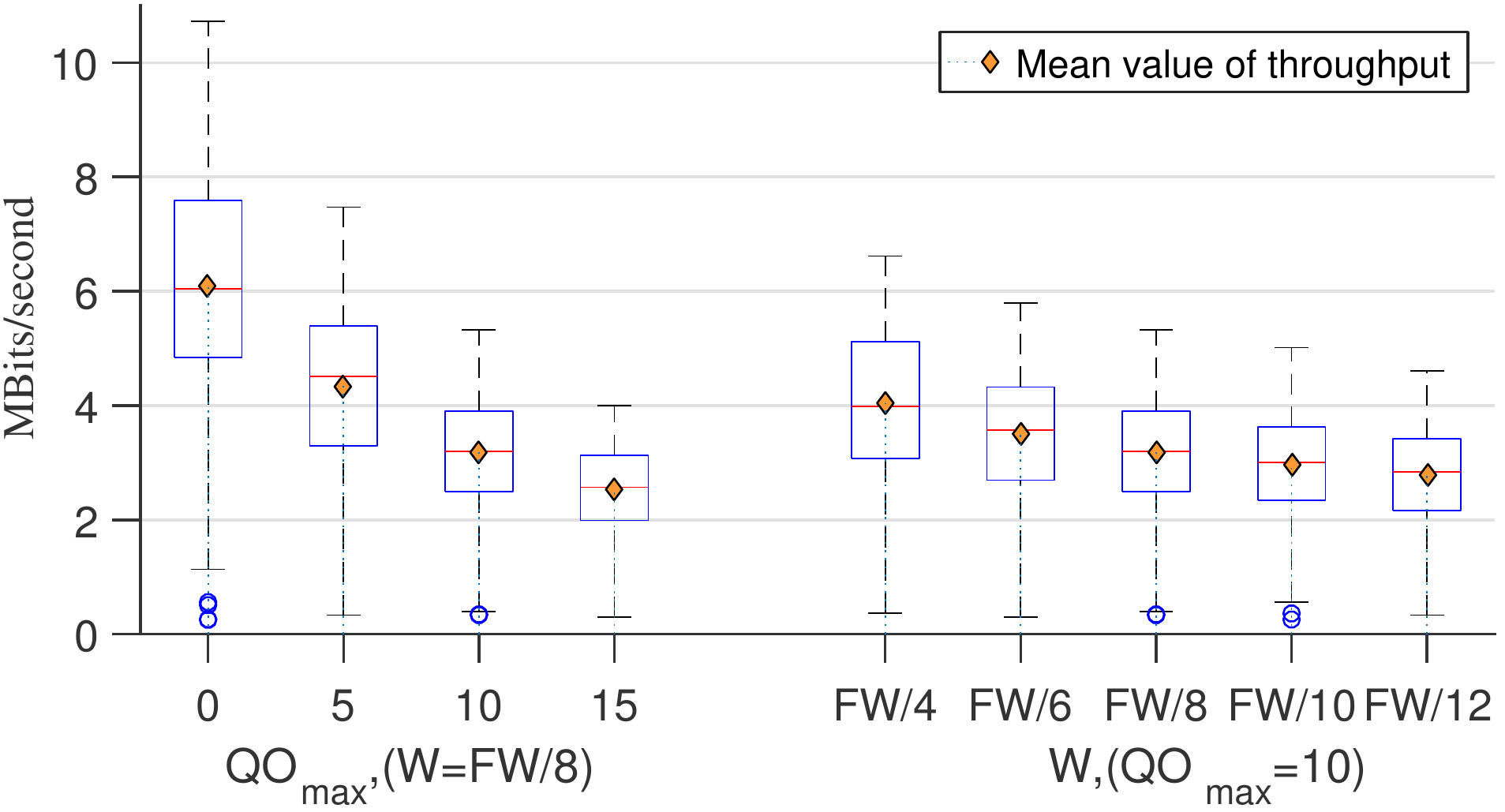}}\\
\subfloat[Little Racers STREET]{\includegraphics[width=0.9\linewidth]{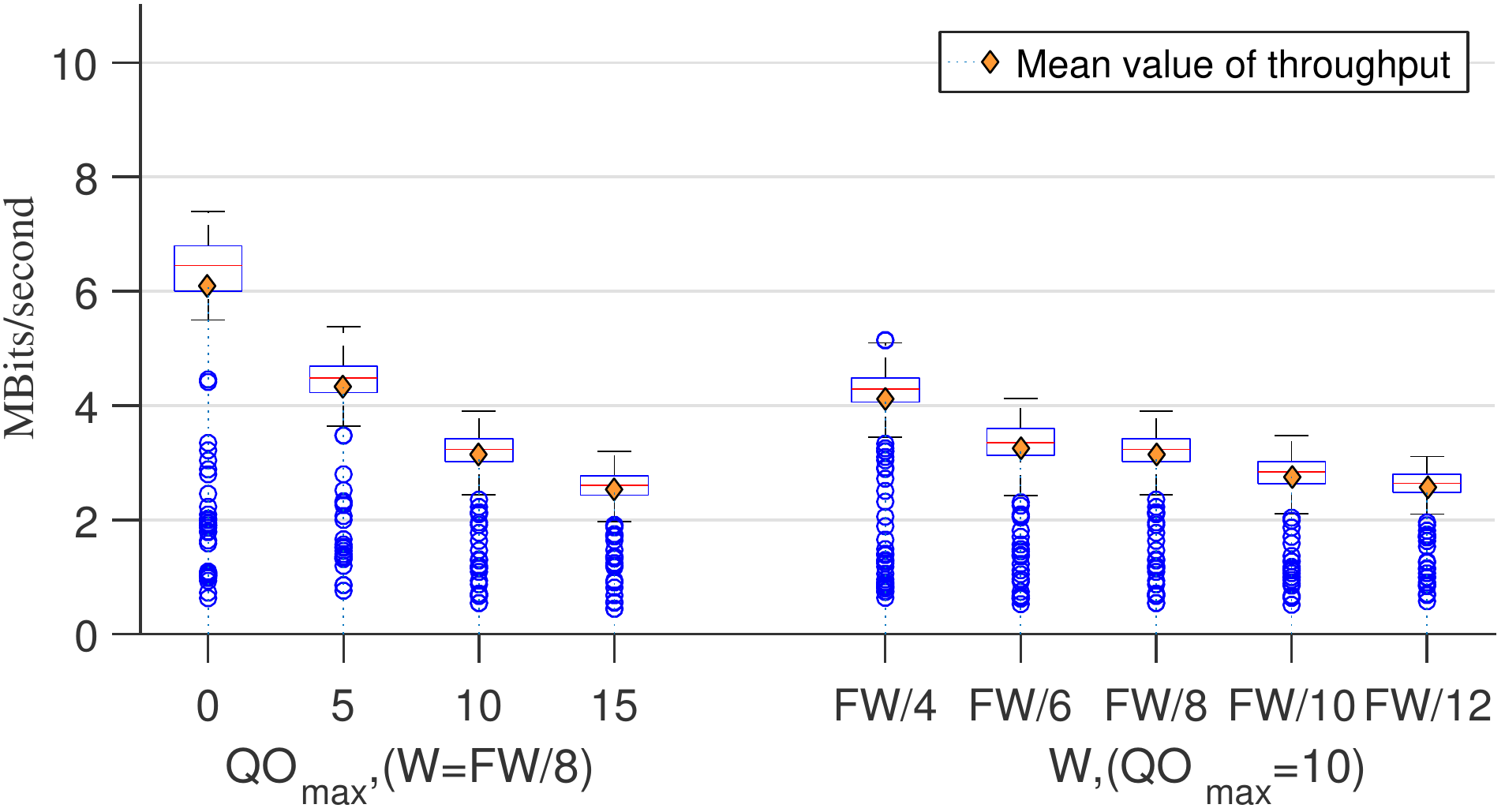}}

\caption{Measured video bitrates with different games and parametrization of foveated streaming. $FW$ is the width of the display in pixels. The box comprises the inter-quartile range, the red line in the middle of the box is the median, and the diamond denotes the mean.}
\label{fig:bitrates}
\end{center}
\end{figure}

Figure \ref{fig:bitrates} shows the results for three games. The first thing to note is that increasing the maximum QP offset dramatically reduces the resulting bitrate. In contrast, decreasing the $W$ parameter value has much less pronounced effect. The reason is that when $W$ takes values of $FW/8$ and smaller, the actual size of the foveal region becomes so small that the number of macroblocks covering it in relation to the total number of macroblocks of the frame is very small. Hence, the additional bandwidth savings by further reducing the size of that region are marginal. $W=FW/8$ seems to provide most of the benefits and smaller foveal regions yield little additional savings.

The second observation is that the differences between the games are small when comparing the average or median bitrates across the different parameter values. However, there are notable differences in bitrate variance. Trine 2 exhibits the most and Little Racers the least variance. In the Little Racers game, there is more or less constant motion but because of the bird's-eye view, there is overall less variation in the graphics. The variance also persists with different parameter values. In summary, while the parametrization has a substantial impact on bandwidth savings, it seems to be game agnostic so that the average bitrates are reduced by similar amounts regardless of the game type.

\begin{figure}[ht]
\begin{center}
\subfloat[$QO_{max}$=0]{\includegraphics[width=0.9\linewidth]{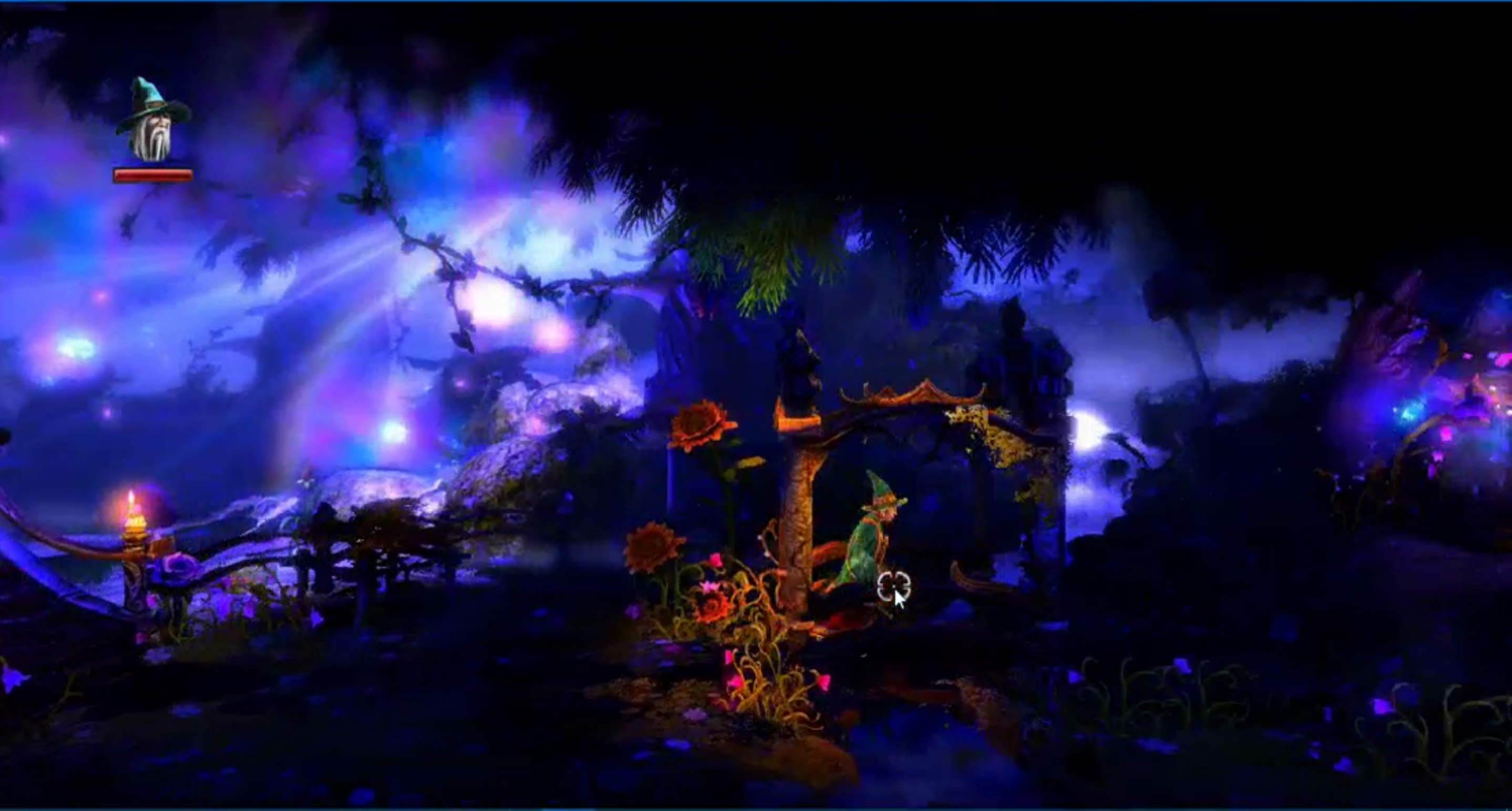}}\\
\subfloat[$QO_{max}$=10,$W$=$FW$/8]{\includegraphics[width=0.9\linewidth]{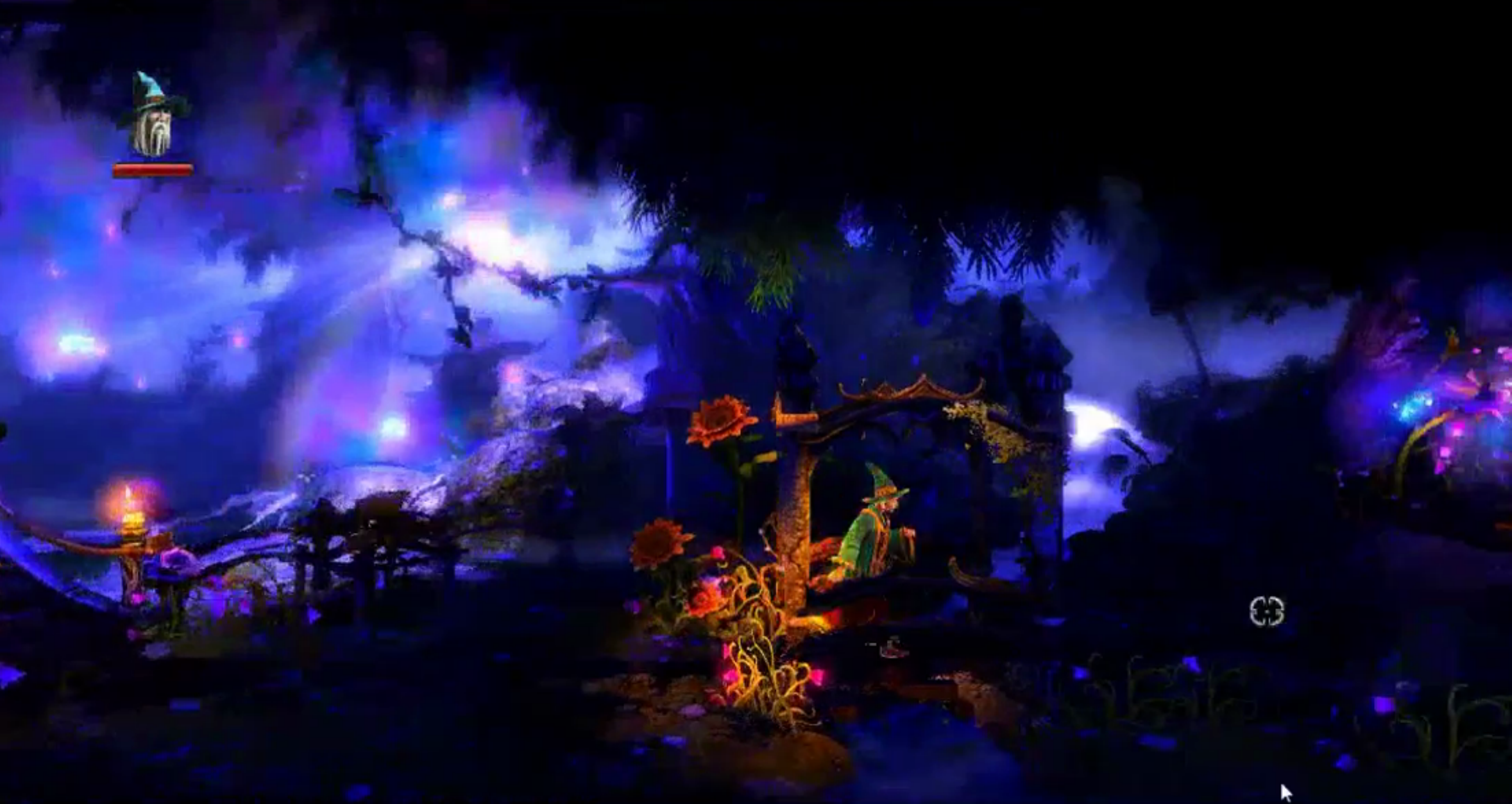}}\\

\caption{Screen Captures of Trine2 with and without offsets. $FW$ is the width of the display in pixels}
\label{fig:screencaps}
\end{center}
\end{figure}
The third observation is that while playing the games using foveated encoding, we did not notice degradation in  quality of experience as long as $QO_{max}$ and $W$ were kept in a reasonable range. Playing different games during our measurements, we noticed that $QO_{max}$=10 and $W$= $FW/8$ provide a reasonably similar quality of experience compared to when using normal encoding. Furthermore, we noticed quality of experience is somewhat less affected in first person shooter games even with  $QO_{max}$=15 and $W$= $FW/8$. This may be due the fact that in such games, the player's gaze is focused on and around the cross-hairs of the weapon which constitutes a  small area of the screen size (see Section \ref{sec:gaze_latency}).  Figure~\ref{fig:screencaps} shows screen captures of the game Trine2 on the cloud gaming client while the player's gaze is located approximately at the running in-game avatar. From visual inspection the image quality at $QO_{max}$=10,  and $W$= $FW/8$, as one focuses on the in game avatar, is nearly the same as that at $QO_{max}$=0. This is somewhat expected as at a normal viewing distance of 50cm (as used in the setup), the area of highest visual acuity has a diameter of approximately 1.8cm. On a laptop screen of width 30cm, this translates to approximately $1/16th$ of the screen width. 
However, quality of experience is highly subjective and a detailed inspection would include user studies which is part of our future work.

\subsection{Gaze Tracking and Latency}
\label{sec:gaze_latency}

\begin{figure}[t]
\begin{center}
\subfloat[AssaultCube]{\includegraphics[width=0.465\linewidth]{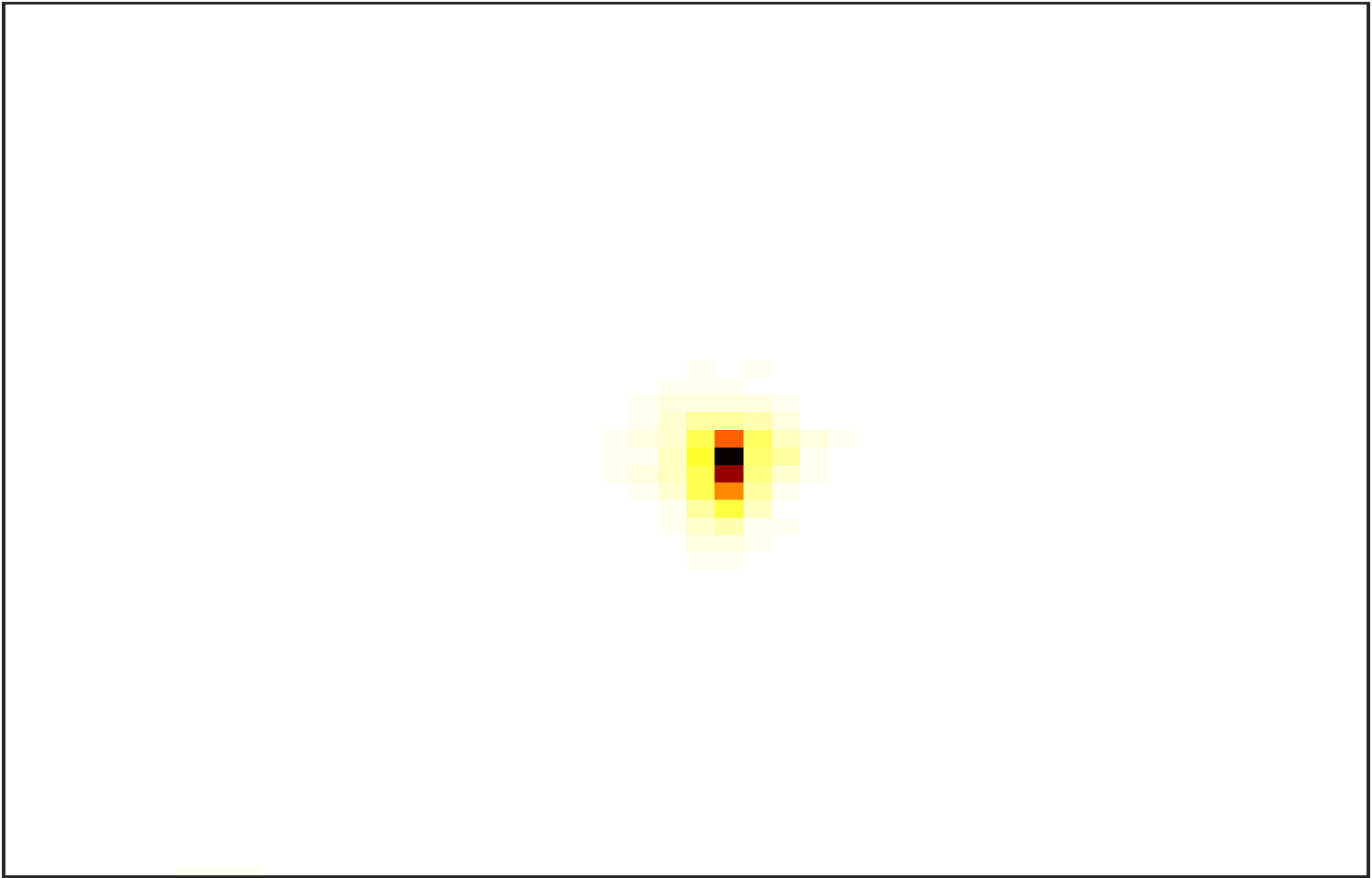}}
\subfloat[Trine 2]{\includegraphics[width=0.49\linewidth]{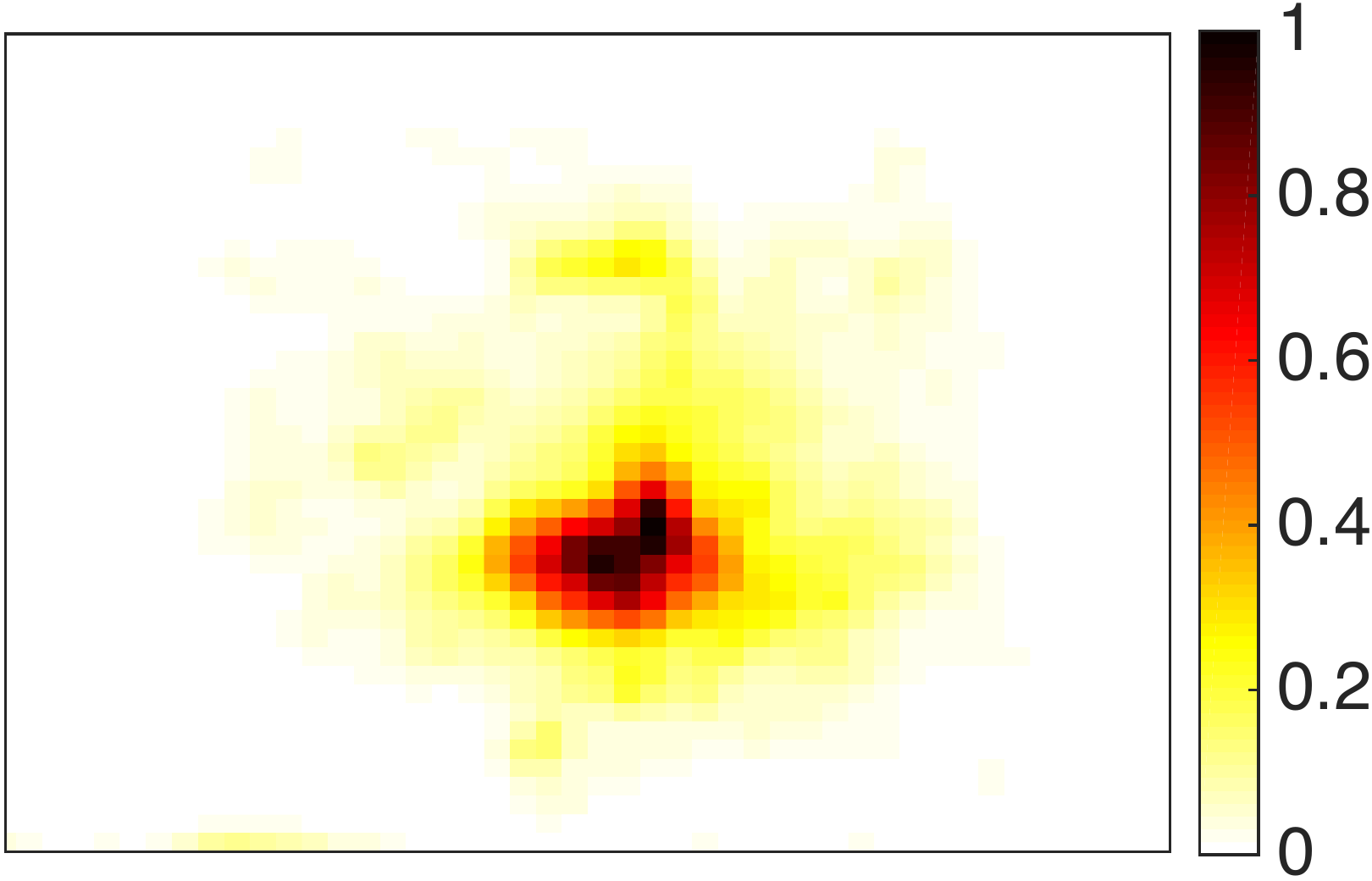}}\\
\subfloat[Little Racers STREET]{\includegraphics[width=0.47\linewidth]{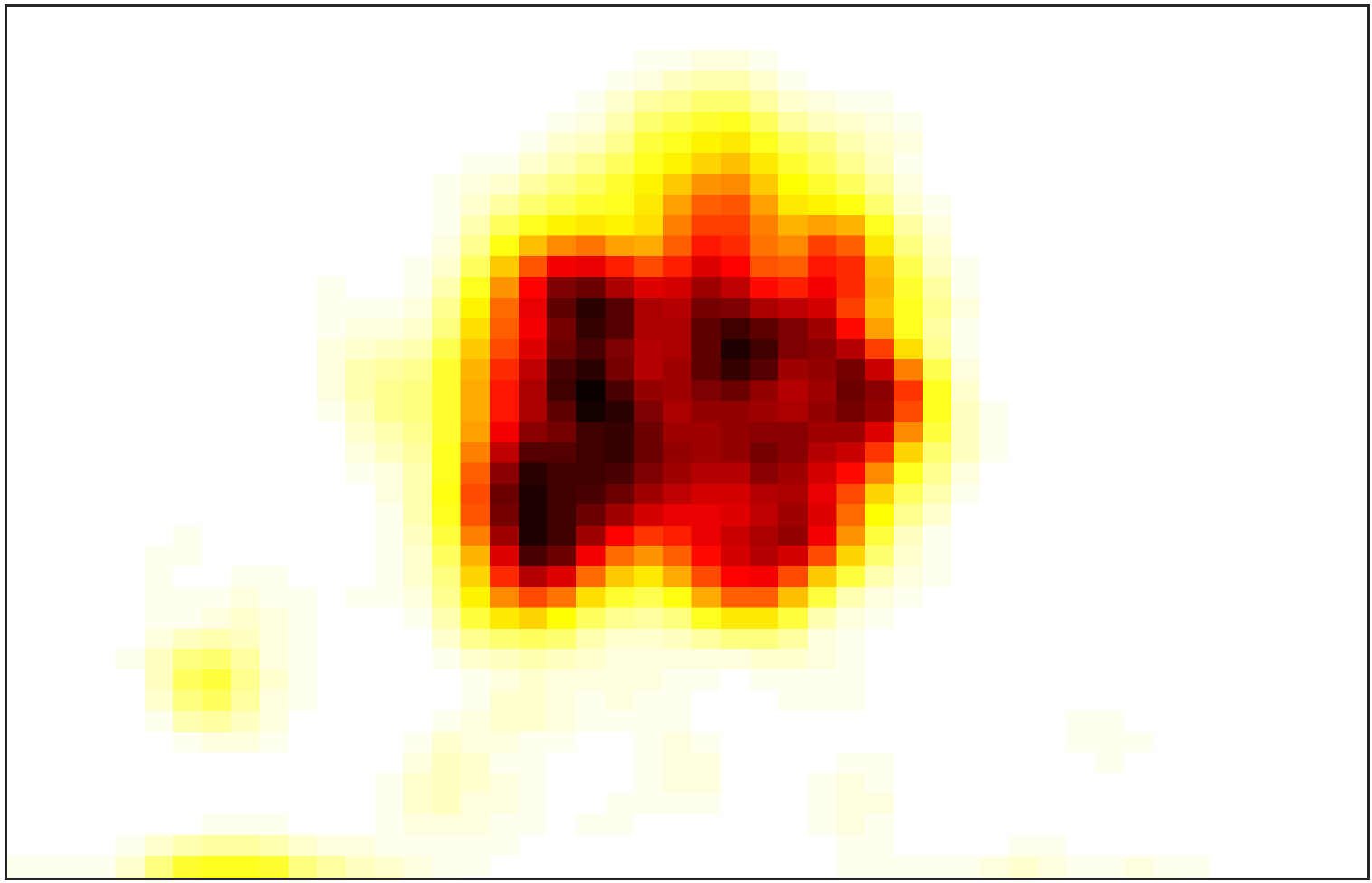}}
\subfloat[Formula Fusion]{\includegraphics[width=0.47\linewidth]{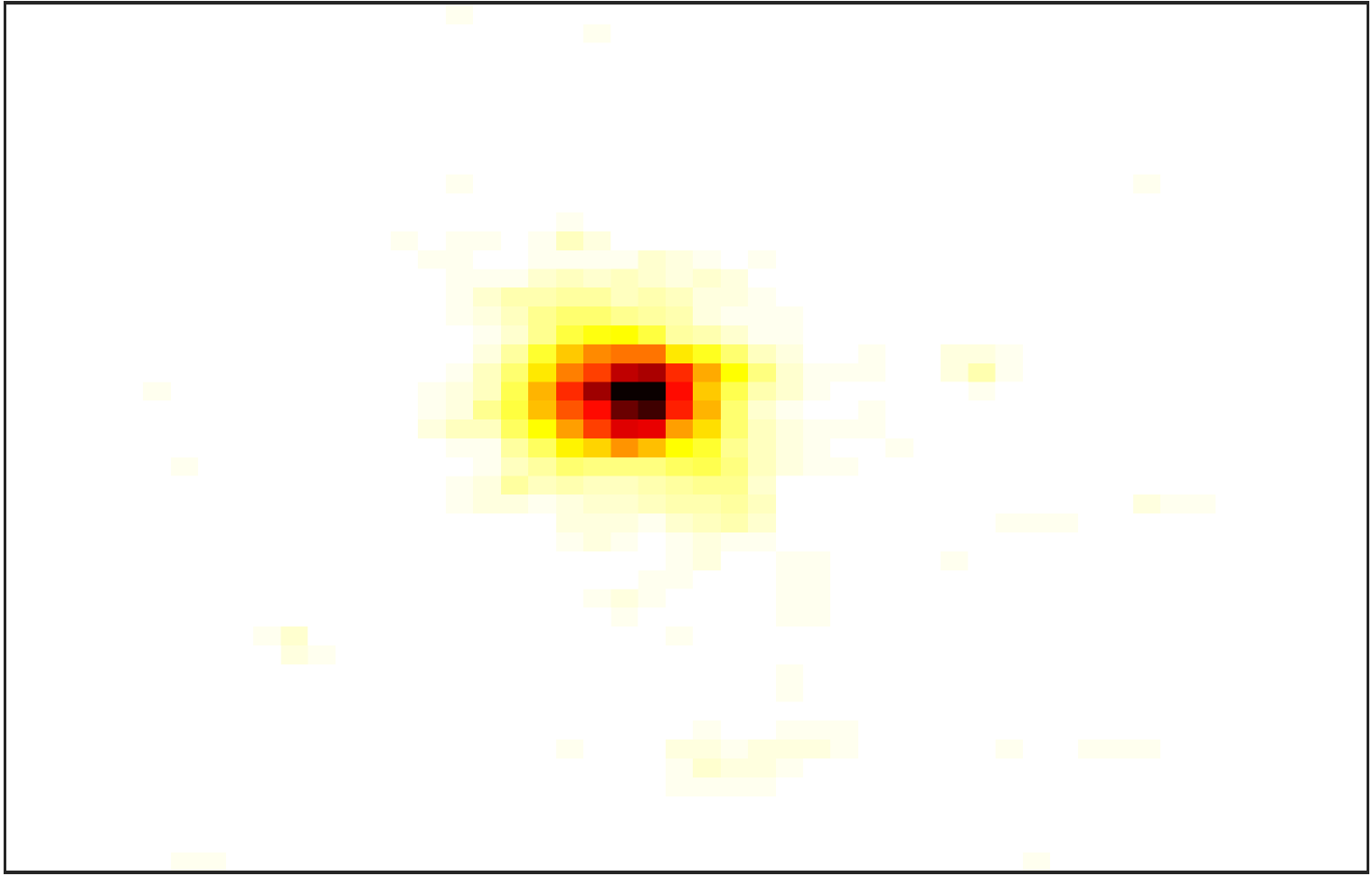}}
\caption{Gaze tracking heatmaps from 15 minute gameplay sessions. The colour scale is normalized.}
\label{fig:gaze_heatmaps}
\end{center}
\end{figure}

Figure \ref{fig:gaze_heatmaps} plots heatmaps of gaze tracking data collected during 15 minute gameplay sessions for four different games. The heatmaps were computed using bivariate Gaussian kernel density estimation applied to gaze coordinates obtained from the Tobii tracker. Occasional glances to peripheral regions are usually so few that they do not show up in the heatmaps. All the heatmaps reveal expectedly that the player mostly looks at the center of the screen but the games differ from each other in terms of the width of the region of player's visual focus. The FPS game AssaultCube keeps the player's gaze focused on a very small region in the center of the display corresponding to the position of the cross hair and the center of what the game character itself "sees". In contrast, Little Racers STREET, the car game controlled from a bird's-eye view, makes the player's gaze wander much more. Consequently, it depends on the game type how challenging it is to provide seamless foveated user experience.

\begin{figure}[t]
\begin{center}
\subfloat[$W=FW/8$]{\includegraphics[width=0.5\linewidth]{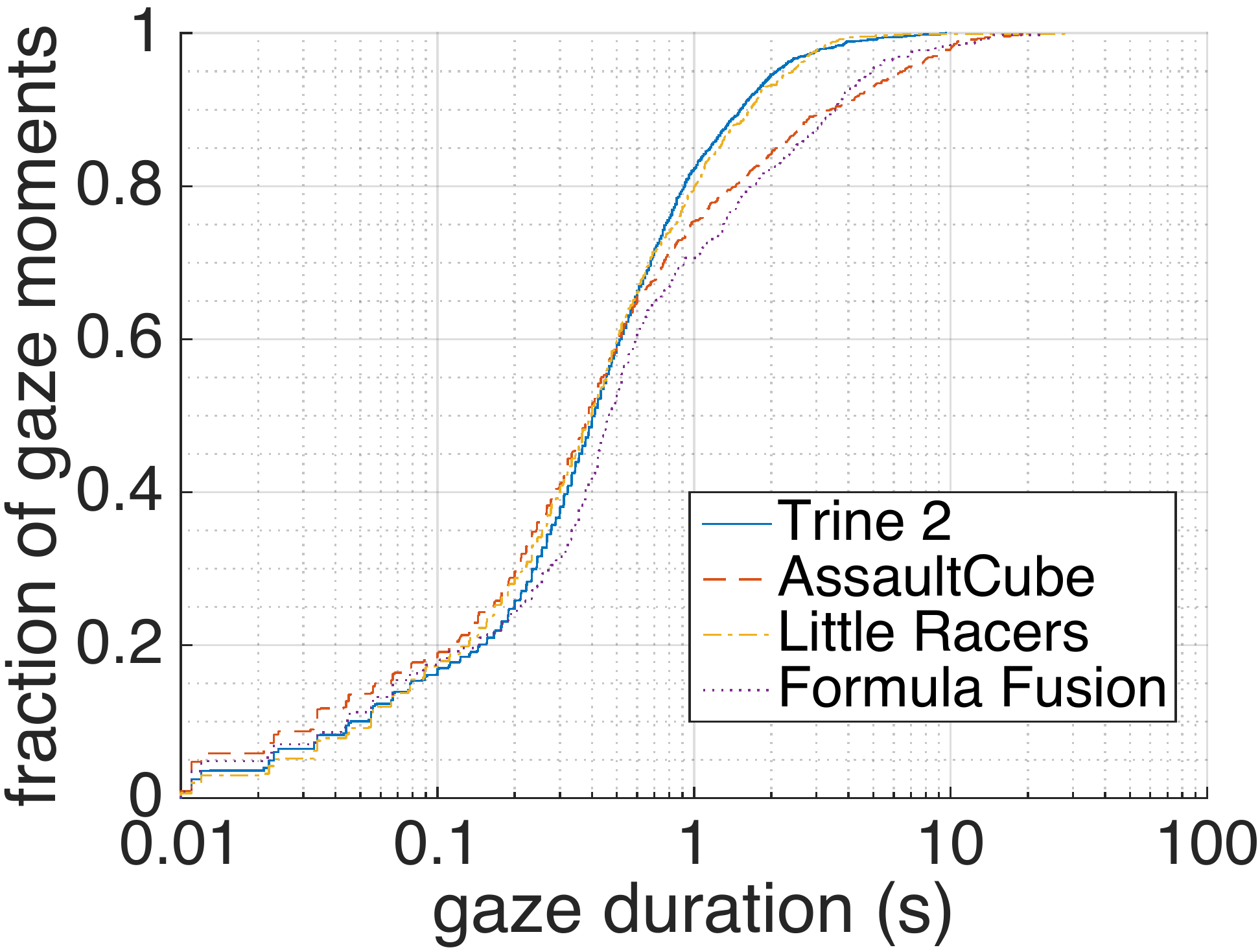}}
\subfloat[$W=FW/4$]{\includegraphics[width=0.5\linewidth]{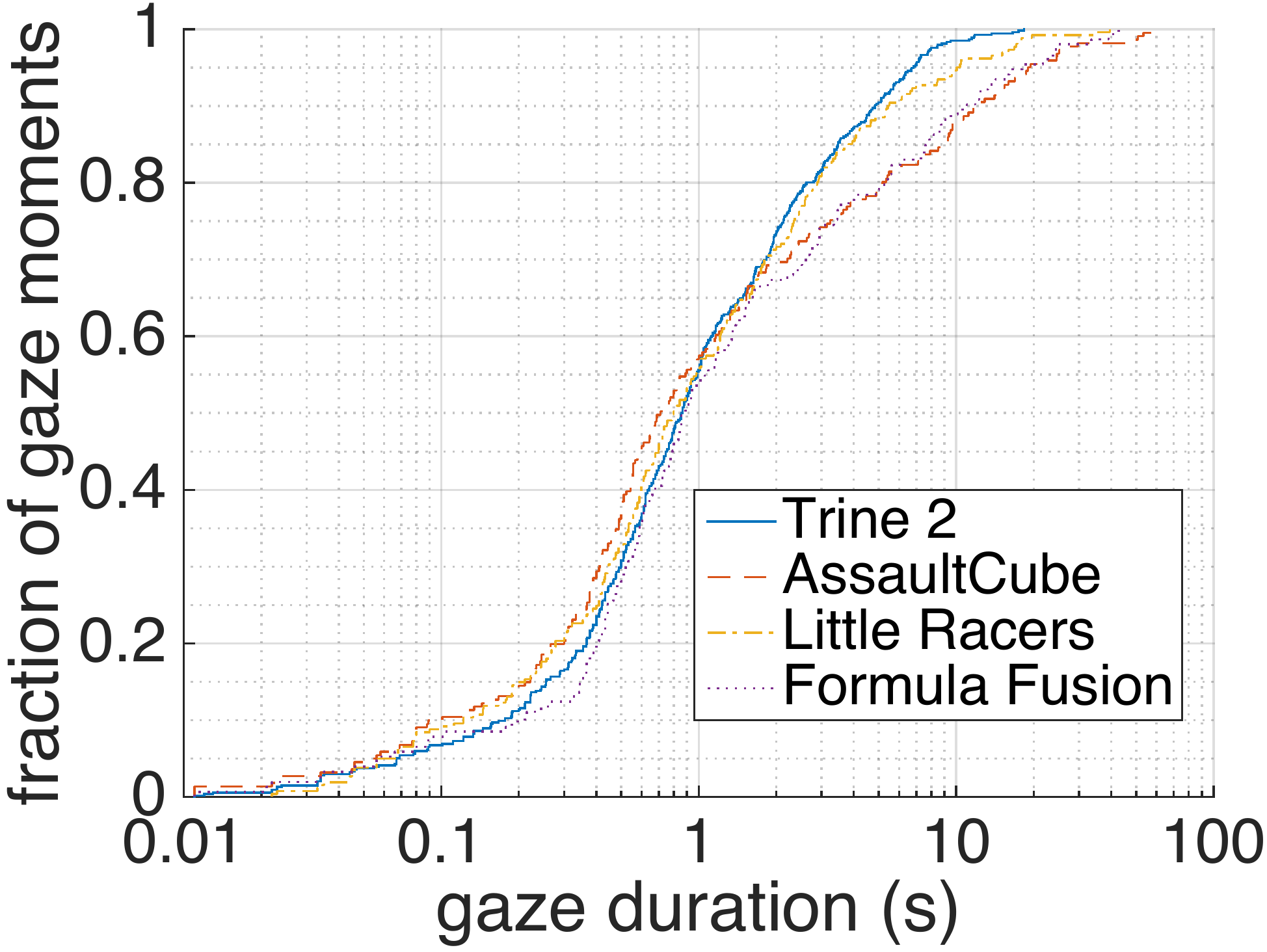}}
\caption{CDF plots of the duration of gaze moments with different sizes of foveated regions.}
\label{fig:gaze_durations}
\end{center}
\end{figure}

Figure \ref{fig:gaze_durations} shows CDF plots of the durations of \textit{gaze moments} which we define as time periods during which the user's gaze does not cross the boundaries of a circular region having a particular size. We observe that there is a very small fraction of gaze moments that last for only one sampling interval (roughly 10 ms). They represent the most difficult scenarios for foveated streaming because of the required latency and may correspond to situations where the player's gaze is rapidly moving across the screen. Roughly 80-90\% of the gaze moments last longer than 100ms and 20-40\% last longer than 1s. AssaultCube and Formula Fusion tend to generate a larger fraction of long gaze moments than the two other games, which agrees with the gaze tracking heatmap results in Figure~\ref{fig:gaze_heatmaps}. 

\begin{figure}[th]
 \begin{center}
   \includegraphics[width=0.7\linewidth]{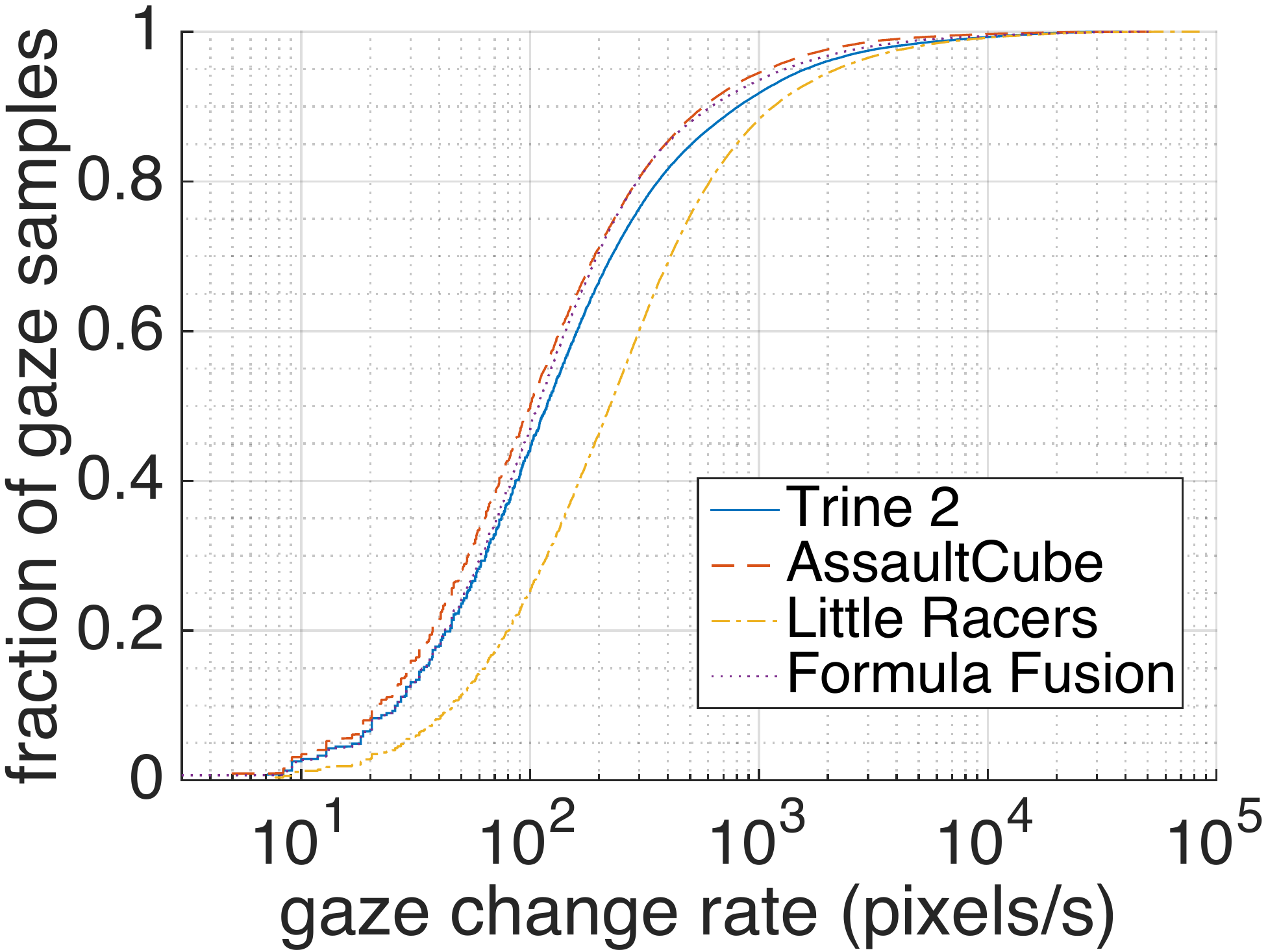}
   \caption{CDF of rate of gaze shifting.}
   \label{fig:gaze_changerate}
 \end{center}
\end{figure}

We also computed the rate of change of gaze during gameplay by dividing the pixel-wise distance of subsequent gaze data samples by the time difference of samples. The CDF of the results is plotted in Figure \ref{fig:gaze_changerate}. There is a notable difference between the Little Racers game and the others so that Little Racers exhibits clearly more rapid gaze shifts (note the logarithmic scale). In general, we observe rarely values beyond 1K px/s which correspond to gaze shifting from one side of the screen to the other within a second\footnote{We should keep in mind that the eye tracker device uses some filtering, although "light filtering", which affects these results but we do not know exactly how much and in which way.}. The framerate in our cloud gaming experiments was 40-45 fps. Hence, over 90\% of the time the gaze would shift less than 25 pixels in between subsequent frames.

Several researchers, including us, have characterized latencies in cloud gaming~\cite{chen2014tmm,kamarainen17mmsys}. Without going into details, having a low latency network path (below 20-30 ms) between the client and server, well provisioned server, and suitable games, it is possible to achieve end-to-end (e2e) latency, i.e., delay from control action to photon, below 100 ms. Even some mobile scenarios can yield below 100 ms e2e latency. The Tobii gaze tracker has a sampling rate of 90Hz, hence it reports gaze positions each 11 ms. To obtain an estimate of the e2e latency for foveation in a cloud gaming setup, i.e. latency from eye-movement to noticeable change on the display, we add the Tobii sampling latency to the e2e cloud gaming latencies\footnote{More precisely, we replace the device-to-kernel part of the control input delay (see measurements in ~\cite{kamarainen17mmsys} for mobile setups) with the gaze tracker's delay. However, that delay is negligible for external USB controllers, so we end up just adding the Tobii delay.}. Comparing the resulting 110 ms to Figure \ref{fig:gaze_durations}, the clear majority of gaze moments should last much longer than the latency of updating the foveated region in the video stream. Figure \ref{fig:gaze_changerate} further suggests that when playing Trine 2, AssaultCube, or Formula Fusion, half of the time the gaze change rate is less than 100 px/s, which means that the gaze shifts at most 11 pixels during the time it takes to turn gaze into foveated video. This seems reasonable given that the effective values for the $W$ parameter, which determines the foveated region, are at least a hundred pixels. While these results are encouraging, we acknowledge that detailed understanding of the impact of this latency on user experience requires subjective studies, which we leave for future work.

\section{Related Work}
\label{sec:related}

Foveated video coding has been studied for over two decades. Wang et al.~\cite{wang2006foveated} wrote a survey on the topic over ten years ago but the techniques have not seen wide scale deployments so far.

Recently, several papers have been published on foveated streaming of precoded video and mostly based on video tiling. D'Acunto et al.~\cite{DAcunto16mmsys} developed a video tiling based solution to allow zooming and navigation within a video streaming. Their solution is based on the MPEG-DASH standard's Spatial Relationship Description (SRD) feature~\cite{niamut16mmsys}. Zare et al.~\cite{zare16mm} use HEVC compliant video tiling method for panoramic video targeting specifically VR scenarios. Similarly, Qian et al.~\cite{qian16atc} study how to selectively stream only the visible parts of panoramic video for clients connected through a mobile network. Ryoo et al.~\cite{ryoo16mmsys} developed a foveated video streaming service based on pre-coding videos using a multi-resolution approach. The gaze tracking of their system is based on webcam input.

The two pieces of work most closely related to ours are by Ahmadi et al.~\cite{ahmadi14msys} and Mohammadi et al.~\cite{mohammadi15icmew}, which both focus on cloud gaming. Ahmadi et al. train a SVM-based attention model offline with eye tracking database and use it in video encoding. Mohammadi et al. propose to use live gaze data like we do but, in contrast to our work, their solution relies on object-based coding and requires modifications to the game engine. Consequently, the solution cannot be applied to and evaluated with off-the-shelf games.


\section{Conclusions and Future Work}
\label{sec:conclusions}

In this work, we proposed to combine foveated streaming with cloud gaming. We developed a prototype system that integrates foveated streaming with off-the-shelf gaze tracker device into state of the art cloud gaming software. Our evaluation results suggest that its potential to reduce bandwidth consumption is significant, as expected. We also demonstrate the impact of different parameter values on the bandwidth consumption with different games and provide some evidence on how to select parameter values. Back of the envelope latency estimations based on related work and gaze tracker specifications combined with gaze data analysis give us reason to be relatively optimistic about the impact on user experience.

As future work, we are planning to examine the quality of experience dimension in more depth through subjective studies. We also intend to investigate the feasibility of mobile cloud gaming with foveated streaming and the possibilities of extending the work towards mobile Virtual Reality.

\section*{Acknowledgment}

This work has been financially supported by the Academy of Finland (grant numbers 278207 and 297892), Tekes - the Finnish Funding Agency for Innovation, and the Nokia Center for Advanced Research.




%
%
%

\end{document}